\documentclass[extra]{gji}
\usepackage{timet}
\usepackage{physics}
\usepackage{graphicx}
\usepackage{float}

\title{Slow periodic oscillation without radiation damping: New evolution laws for rate and state friction}
\author[Mizushima and Hatano]{Ryo Mizushima and Takahiro Hatano \\
	Department of Earth and Space Science, Osaka University, 560-0043 Toyonaka, Japan
}
\date{}
\pagerange{\pageref{firstpage}--\pageref{lastpage}}
\volume{229}
\pubyear{2022}

\newtheorem{theorem}{Theorem}[section]
\begin{document}
	\label{firstpage}
	\maketitle
	
	\begin{summary}
		The dynamics of sliding friction is mainly governed by the frictional force. Previous studies have shown that the laboratory-scale friction is well described by an empirical law stated in terms of the slip velocity and the state variable. The state variable represents the detailed physicochemical state of the sliding interface. Despite some theoretical attempts to derive this friction law, there has been no unique equation for time evolution of the state variable. Major equations known to date have their own merits and drawbacks. To shed light on this problem from a new aspect, here we investigate the feasibility of periodic motion without the help of radiation damping. Assuming a patch on which the slip velocity is perturbed from the rest of the sliding interface, we prove analytically that three major evolution laws fail to reproduce stable periodic motion without radiation damping. Furthermore, we propose two new evolution equations that can produce stable periodic motion without radiation damping. These two equations are scrutinized from the viewpoint of experimental validity and the relevance to slow earthquakes.
	\end{summary}

	\section{Introduction}
	\subsection{Evolution equations for rate- and state-dependent friction law}
	Since slip behaviours of earthquake faults are mainly dominated by the frictional force, study of friction laws plays a vital role in understanding earthquakes from physical point of view. Data from laboratory experiments on quasi-static slip and isothermal condition of rocks are well summarized into an empirical law, which is referred to as the rate- and state-dependent friction law (hereafter the RSF law) \citep{Dietrich1979}.
	\begin{equation}
	\label{RSF}
		\mu = \mu_{0} + a\log(\frac{\hat{V}}{V_{0}})+b\log(\frac{V_{0}\hat{\theta}}{L}),
	\end{equation}
	where $\hat{V}$ is the slip velocity, $\hat{\theta}$ is the state variable, $L$ is the characteristic length, and $V_{0}$ and $\mu_{0}$ are constants involving an arbitrary reference state.	Two nondimensional parameters $a$ and $b$ depend on materials and various experimental condition. The logarithmic nature of this law reflects a thermal activation process at asperities \citep{Heslot1994}.
	In view of mathematical convenience, we adopt the natural logarithm unless otherwise indicated.
	
	
	The state variable $\theta$ has the dimension of time representing the condition of the slipping surface.
	The behaviour of the state variable is described by a time evolution equation.
	\begin{equation}
	\label{time_evolution}
		\dv{\hat{\theta}}{\hat{t}} = f(\vb{X}(t)),
	\end{equation}
	where $\vb{X}(t)$ denotes relevant macroscopic variables such as the slip velocity, state variable(s), or the shear stress.
	We then need to choose an appropriate function for the right hand side.
	However, since there is no rigorous definition of the state variable, derivation of a time evolution equation cannot be systematic but only heuristic.
	As a result, there has been no decisive form for the function $f$ in Eq. (\ref{time_evolution}).

	Here we briefly review some of the most commonly used equations \citep{Ruina1983}:
	\begin{eqnarray}
	\label{sliplaw}
		\dv{\hat{\theta}}{\hat{t}} &=& -\frac{\hat{V}\hat{\theta}}{L}\log(\frac{\hat{V}\hat{\theta}}{L}),\\
		\dv{\hat{\theta}}{\hat{t}} &=& 1-\frac{\hat{V}\hat{\theta}}{L}.
	\end{eqnarray}
	The former equation, referred to as the slip law, leads to an exponential relaxation of the friction force after an abrupt velocity change.
    The latter equation is referred to as the aging law. It describes monotonic increase of the state variable, which is interpreted as time-dependent healing or aging. The slip law does not describe such a healing effect. These laws have their own merits and drawbacks, but somewhat describe experimental results under isothermal and low-velocity conditions.
    
	Less common but more recent equation is the following \citep{Nagata2012}.
	\begin{equation}
	\label{nagata}
		\dv{\hat{\theta}}{\hat{t}}=1-\frac{\hat{V}\hat{\theta}}{L}-\frac{\gamma}{b\sigma}\dv{\hat{\tau}}{\hat{t}} \hat{\theta},
	\end{equation}
	where $\gamma$ is a positive constant, $\sigma$ is the normal stress, and $\tau$ is the shear stress.
	

	These three evolution laws have their own merits and drawbacks \citep{Marone1998,Bizzarri2011}.
	At a steady state, however, each equation gives $\hat{\theta} = L/\hat{V}$, leading to
	\begin{equation}
		\label{muss}
		\mu=\mu_{0}+(b-a)\log(\frac{\hat{V}}{V_{0}}).
	\end{equation}

	\subsection{Feasibility of periodic motion with RSF law}
	Using the RSF law as a boundary condition, the motion of faults is defined and solved numerically based on the elasticity theory \citep{Cochard1994}.
	Linear stability analysis under some simple configurations (e.g., anti-plane shear) reveals that uniform and steady slip can be unstable via a Hopf bifurcation, where long-wavelength perturbations grow exponentially with time \citep{RiceRuina1983,Rice2001}.
	
	To characterize the oscillatory motion due to Hopf bifurcation, linear stability analysis alone is not sufficient since the oscillation is mainly governed by nonlinear effects in the friction force. To this end, nonlinear analysis must be implemented \citep{Gu1984,Ranjith1999,Viesca2016,Viesca2016PRE}. In particular, for a single degree of freedom system, \cite{Gu1984} and \cite{Ranjith1999} construct quantities similar to a Lyapunov function and show rigorously that any trajectory moves away from the unstable fixed point. Their results also imply that the limit cycle is absent and the trajectory diverges. This is rather a major drawback of existing time evolution equations for the RSF. This has been already pointed out by \cite{Rice1993}, but does not appear to be commonly recognized.

	To prevent the divergence of trajectory and to yield the stable cycle of oscillation, some extra ingredients have been introduced. For instance, if inertia is introduced to the equation, it yields a steady oscillatory motion. However, the equation with inertia corresponds to the motion of a rigid body pulled with an elastic spring \citep{Sugiura2012}, but cannot be a model for a slip patch.
	More commonly used technique is to introduce a viscous resistance that is proportional to the slip velocity. This resistance is called as the "radiation damping" \citep{Rice1993}, which can be derived from the dynamic elasticity theory \citep{Cochard1994}.
	A popular model for slow earthquakes, which is referred to as a quasi-static (or quasi-dynamic) model, assumes the static Green's function together with the radiation damping. It may be, however, an {\it ad hoc} hybrid of dynamic/static elasticity. More importantly, use of radiation damping in modelling slow aseismic events may be somewhat contradictory since they do not radiate seismic waves.
	
    Some studies have shown that, if the friction is velocity strengthening at higher velocities, aseismic slow oscillation can be realized \citep{Matsuzawa2010, Shibazaki2007, Barbot2019}. However, these models also adopt radiation damping. To the best of our knowledge, there have been very few studies that reproduce aseismic slip cycles without radiation damping \citep{Hawthorne2013}. However, the model settings in \cite{Hawthorne2013} appear rather special in terms of the spatial distribution of frictional parameters and the boundary conditions. It is thus not clear how general such a model can reproduce an aseismic slip cycle without radiation damping. Some models may be able to reproduce transitory cycles of aseismic slips, but such a transitory cycle is not our aim here. Rather, we wish to reproduce a steady aseismic cycle in the sense that it is repeatable for arbitrary times. 
   
	In this paper, we shed new light on this problem.
	Using a mathematical theorem, we show analytically that the limit cycle does not exist in a single degree of freedom system with popular evolution equations. We then propose new evolution equations and show that they reproduce periodic motion repeatedly within the time of simulation (more than one hundred times). We also show that, in contrast to some previous studies, velocity-strengthening is not an essential ingredient to realize a slow aseismic slip.
	In Section 2, we formulate the problem by introducing the fault patch model and review some previous results. In Section 3, we prove that three major evolution laws cannot yield any periodic motion. In Section 4, we propose two new evolution laws that can realize periodic motion. We then discuss the behaviours of these new equations in the context of velocity step experiments and examine the validity of these laws. In Section 5, we discuss the relevance of these new equations in the context of slow earthquakes \citep{Obara2016}.
	
	\section{Model}
	\subsection{Single degree of freedom system}
	Since nonlinear analysis of partial differential equations is not straightforward, slip dynamics on a spatially extended system is often reduced to a single degree of freedom system \citep{Scholz}. This is done by considering that the slip velocity on a fault plane is perturbed only in a finite circular area, and the rest of the fault plane is displaced at a constant slip velocity, $V_{p}$. The relevant variables are then the perturbed slip velocity $\dot{\hat{u}}$ and the displacement ${\hat{u}}$ at the patch center.
	
	The shear stress $\tau$ on the patch is proportional to the perturbed displacement with the effective stiffness $\hat k$: $\tau= \hat{k}(V_{p}\hat{t}-\hat{u})$.
	Since the applied shear stress and frictional force must balance on the fault plane, the following equation holds:
	\begin{equation}
		\label{balance}
		\frac{\hat{k}}{\sigma}(V_{p}\hat{t}-\hat{u}) = \mu_{0} 
		+ a\log(\frac{\hat V}{V_{0}})+b\log(\frac{V_{0}\hat{\theta}}{L}),
	\end{equation}
	where $\hat V =d \hat u/dt$ and $\sigma$ is the normal stress.

	\subsection{Definition of dimensionless variables}
	Hereafter we use dimensionless variables given as follows
	\begin{equation}
		\label{nondim}
		u = \frac{\hat u}{L}, \, V=\frac{\hat{V}}{V_{p}}, \,  \theta=\frac{\hat{\theta}V_{p}}{L}
		, \, k=\frac{\hat{k}L}{\sigma}  , \, t=\frac{\hat{t}V_{p}}{L}.
	\end{equation}
	Here, $V_{p}$, $k$, and $L$ are the steady-state slip velocity, the spring constant, and the characteristic length, respectively.
	
	\subsection{Governing equations}
	Assuming that the velocity and the state variable change smoothly, we may differentiate the equation with respect to time. Using (\ref{nondim}), we can derive the nondimensional equations.
	\begin{equation}
		\label{patch}
		\dv{V}{t}=\frac{k}{a}(1-V)V-\frac{b}{a\theta} f(V, \theta)V,
	\end{equation}
	\begin{equation}
		\label{evolve}
		\dv{\theta}{t}=f(V, \theta).
	\end{equation}
	After differentiation, the system is reduced to an autonomous system with two variables, making the analysis easier. 
	
	A linear stability analysis of this system reveals that a Hopf bifurcation occurs when the stiffness $k$ is smaller than the critical value $k_{c}$ \citep{Ruina1983}. However, in normal linear stability analysis, one can know the flow field only in a closed proximity of the fixed point. To know the solution realized in the linearly unstable regime, one needs to know the global flow field.
	
	\section{Absence of periodic motion: proof}
	Here we scrutinize three major evolution equations as to whether they can yield stable periodic solutions.
	
	\subsection{Dulac's criterion and its implication}
	In two dimensional dynamical systems, some useful theorems guarantee the absence of closed orbits. Here we make use of the following Dulac's criterion \citep{Strogatz2001}. The proof is described in Supplemental Information.
	
	\theorem{Dulac's criterion} \\
	The vector $\vb x$ is defined in a set $A$, which is a simply connected area on $R ^ {2}$. Time evolution of the vector $\vb x$ is described by $\vb{\dot{x}}=\vb{f}(\vb{x})$, where $\vb{f}(\cdot)$ is a differentiable vector field. Let $g(\vb{x})$ be a real-valued function. If the sign of $\nabla \vdot({g \vb{\dot{x}}})$, either positive or negative, is the same at any point on $A$, then there is no closed orbit.
	\endtheorem
	
	If there is no closed orbit, a limit cycle cannot exit. We can thus prove the absence of limit cycle by making use of the above theorem. In the present context, the vector field $\vb{\dot{x}}$ is given by Eqs. (\ref{patch}) and (\ref{evolve}). Then, to prove the absence of the limit cycle, we need to find a function $g(\cdot)$ that satisfies the criterion.
	
	Since the present system is described by two variables, the trajectory is limited to the two-dimensional space, $(V,\theta)$. According to the theory of dynamical systems \citep{Strogatz2001}, the asymptotic behavior of two-dimensional systems is limited to three cases: fixed point, limit cycle, and divergence to infinity. (Note that strange attractor is ruled out in two dimensional space). Therefore, if the fixed point becomes unstable and the limit cycle does not exist, the trajectory must diverge after a sufficiently long time. This holds for all the three cases below.
	
	Note that, since $V>0$, it is sufficient to consider the first quadrant in $(V,\theta)$ space; i.e., $V>0$ and $\theta>0$.
	
	\subsection{Case 1: Slip law}
	In the case of slip law, the governing equations are
	\begin{equation}
		\dv{V}{t} =\frac{k}{a}(1-V)V+\frac{bV^2}{a}\log(V\theta),
	\end{equation}
	\begin{equation}
		\dv{\theta}{t} =-V \theta \log(V\theta).
	\end{equation}
	This system has a unique fixed point at $(V, \theta)=(1, 1)$. Jacobi matrix J is then derived as
	\begin{equation}
		J \equiv \left(
		\begin{array}{cc}
			\displaystyle\pdv{\dot{V}}{V} & \displaystyle\pdv{\dot{V}}{\theta} \\
			\displaystyle\pdv{\dot{\theta}}{V} & \displaystyle\pdv{\dot{\theta}}{\theta} \\
		\end{array}
		\right)_{(V, \theta)=(1, 1)}
		= \left(
		\begin{array}{cc}
			\displaystyle{\frac{b-k}{a}}& \displaystyle{\frac{b}{a}} \\
			-1 & -1\\
		\end{array}
		\right).
	\end{equation}
	Therefore the eigenvalues are
	\begin{equation}
		\lambda = \frac{1}{2}\left[\frac{b-a-k}{a}\pm\sqrt{\left(\frac{b-a-k}{a}\right)^{2}-4\frac{k}{a}}\right].
	\end{equation}
	Thus, if the spring constant $k$ is smaller than the critical value $k_{c}=b-a$, $\Re{\lambda}$ and $\Re{\bar{\lambda}}$ are positive and the fixed point is unstable. In other words, Hopf bifurcation occurs at $k=k_{c}$, below which steady slip is unstable.
	
	We then prove the absence of limit cycle in this system based on the Dulac's criterion.
	By choosing 
	\begin{equation}
	g(V,\theta) =\frac{a}{V}\theta^{(b/a-1)}, 
	\end{equation}
	the divergence of $g\vb{\dot{x}}$ is calculated as
	\begin{equation}
		\label{slip div}
		\nabla \vdot({g\vb{\dot{x}}}) = \theta^{(\frac{b}{a}-1)}(b-a-k)=\theta^{(\frac{b}{a}-1)}\epsilon,
	\end{equation}
	where $\epsilon$ is a constant defined as $\epsilon=k_{c}-k$. The divergence Eq. (\ref{slip div}) is always positive except at the bifurcation point, $k=k_c$.
	Namely, there is no limit cycle if $k<k_{c}$.
	
	\subsection{Case 2: Aging law}
	In the case of aging law, the governing equations are
	\begin{equation}
		\dv{V}{t}=\frac{k}{a}(1-V)V-\frac{b}{a\theta}(1-V\theta)V,
	\end{equation}
	\begin{equation}
		\dv{\theta}{t}=1-V\theta.
	\end{equation}
	This system also has a unique fixed point at $(V,\theta)=(1, 1)$. Jacobi matrix J is then derived as
	\begin{equation}
		J = \left(
		\begin{array}{cc}
			\displaystyle{\frac{b-k}{a}} & \displaystyle{\frac{b}{a}} \\
			-1 & -1\\
		\end{array}
		\right)
	\end{equation}
	Then we calculate the eigenvalues of J and find the Hopf bifurcation point as $k_{c}=b-a$.
	
	We can again prove the absence of limit cycle in this model by setting 
	\begin{equation}
	g(V,\theta)=\frac{a}{V}.
	\end{equation}
	This leads to
	\begin{equation}
		\label{aging div}
		\nabla \vdot({g\vb{\dot{x}}}) = -k+b-a = \epsilon,
	\end{equation}
	where $\epsilon$ is constant defined as $\epsilon=k_{c}-k$. The divergence Eq. (\ref{aging div}) is always positive for $k<k_c$. This result means that there is no limit cycle once the fixed point is unstable.
	
	\subsection{Case 3: Nagata's law}
	In the case of Nagata's law, the governing equations are
	\begin{equation}
		\dv{V}{t}=\frac{k}{a}(1-V)V - \frac{b}{a\theta}\left[1-V\theta-\frac{\gamma k}{b}(1-V)\theta\right]V,
	\end{equation}
	\begin{equation}
		\dv{\theta}{t} = 1 -V\theta -\frac{\gamma k}{b}(1-V)\theta.
	\end{equation}
	This system has a unique fixed point at  $(V, \theta)=(1, 1)$. At this point the Jacobi matrix is
	\begin{equation}
		J = \left(
		\begin{array}{cc}
			\displaystyle{\frac{-k+b-\gamma k}{a}} & \displaystyle{\frac{b}{a}} \\
			\displaystyle -1+\frac{\gamma k}{b} & -1
		\end{array}
		\right)
	\end{equation}
	Then we calculate the eigenvalues of J and find that Hopf bifurcation occurs at $ k_{c} = (b-a)/(\gamma +1)$.
	
	We then prove the absence of limit cycle in this system by choosing the function $g$ as
	\begin{equation}
	g(V,\theta) = \frac{a}{V}(V^{\frac{a}{b}}\theta)^{\gamma}.
	\end{equation}
	After some calculation, one gets
	\begin{equation}
		\label{Nagata div}
		\div({g\vb{\dot{x}}}) =\theta^{\gamma}V^{\frac{a\gamma}{b}}(b-a-(1+\gamma)k) = \epsilon (1+\gamma)(\theta V^{\frac{a}{b}})^{\gamma}.
	\end{equation}
	This quantity is always positive if the system is unstable ($\epsilon>0$).
	Therefore, the limit cycle is again not feasible for this system.
	
	\section{Proposal of new evolution laws}
	In this section, we propose two novel evolution equations and demonstrate that they yield the limit cycle and thus stable periodic oscillation. Note that radiation damping is not necessary for these new equations.
	
	To study the nonlinear nature of the solution, we make use of the normal form theory and derive the first Lyapunov coefficient, which is denoted by $l_{1}$. This is the function given by the coefficients of the cubic term when the dynamical system is expanded around the Hopf bifurcation point. The sign of this function determines the type of Hopf bifurcation. If $l_{1}$ is positive, then the bifurcation is supercritical. In this case, it is mathematically guaranteed that there is a limit cycle in the vicinity of the fixed point. The details are described in Appendix B.
	
	\subsection{Modified ageing law I}
	We first show that a simple improvement of the aging law can prevent the divergence of trajectory. The first equation we wish to propose is the following.
	\begin{equation}
		\label{power}
		\dv{\hat{\theta}}{\hat{t}}=1-\left(\frac{\hat{V}\hat{\theta}}{L}\right)^{n},
	\end{equation}
	where $n$ is a positive constant. At steady states, this equation yields a simple logarithmic dependence on the slip velocity, Eq. (\ref{muss}).
	Note that Eq. (\ref{power}) with $n=1$ is reduced to the ageing law, the evolution equation Eq. (\ref{power}) is regarded as an extension of the ageing law for $n\neq 1$.
	
	Dimensionless form of the single degree of freedom model then reads
	\begin{equation}
		\label{power1}
		\dv{V}{t}=\frac{k}{a}V(1-V)-\frac{b}{a\theta}\left[1-(V\theta)^{n}\right]V,
	\end{equation}
	\begin{equation}
		\label{power2}
		\dv{\theta}{t}=1-(V\theta)^{n}.
	\end{equation}
	This system has the fixed point: $(V, \theta)=(1, 1)$. Jacobi matrix J at this point is
	\begin{equation}
		J=\left(
		\begin{array}{rr}
			\displaystyle{\frac{-k+nb}{a}} & \displaystyle{\frac{nb}{a}} \\
			-n & -n \\
		\end{array}
		\right).
	\end{equation}
	By calculating the eigenvalues, we obtain the critical stiffness at which Hopf bifurcation occurs.
	\begin{equation}
	\label{kcI}
	k_{c}=n(b-a),
	\end{equation}
	where $n>0$ and $b>a$ are required.
	For instance, in the case of $n=1/2$, supercritical Hopf bifurcation occurs at $k_{c}=(b-a)/2$. 

	We then calculate the first Lyapunov coefficient by using the normal form theory.
	\begin{equation}
		l_{1}(0) = \frac{1}{4}\left(\frac{b-a}{a}\right)^\frac{3}{2}(n-1)
	\end{equation}
	Note that the absolute value may differ depending on how the critical eigenvectors are normalized. Since $b>a$ and $n>0$, $l_{1}(0)$ is negative if $0<n<1$.
	It means that the bifurcation is supercritical and the limit cycle exists.
	
\begin{figure}
	\includegraphics[width=\linewidth]{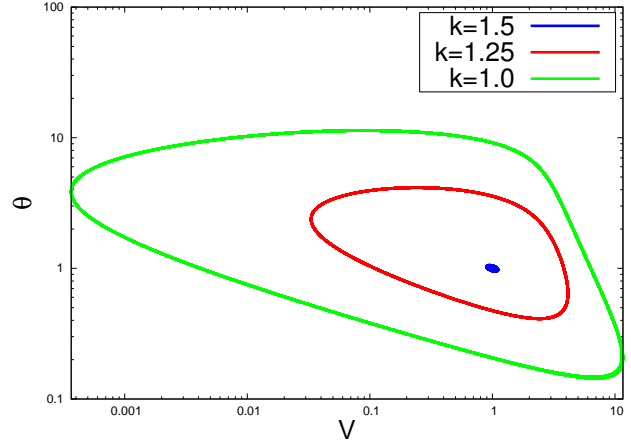}
	\caption{Limit cycles produced by Eqs. (\ref{power1}) and (\ref{power2}). Parameters are set to $n=1/2$, $b=4.0$ and $a=1.0$, so $k_{c}=1.5$. Calculations are conducted for $k=1.5$, $1.25$ and $1.0$.}
	\label{cycle_A}
	\end{figure}

	The numerical result is shown in Figure \ref{cycle_A}, where the limit cycle exists and the amplitude increases as $k$ decreases. Therefore, we confirm a supercritical Hopf bifurcation. More importantly, a periodic motion is realized without introducing radiation damping.

	We also inspect the transient behaviour by considering the velocity step test. Here the velocity is switched between $V_p$ and $10V_p$.
	In this case, it is more convenient to use $V_p$ in place of an arbitrary constant $V_0$ in the RSF, i.e., Eq. (\ref{RSF}).
	We then introduce
	\begin{equation}
		\label{deltamu}
		\delta \mu = \mu-\mu_{0}=a\log V+b\log \theta
	\end{equation}
	with the dimensionless form of the evolution equation.
	\begin{equation}
		\label{nondimpow}
		\dv{\theta}{t}=1-(V\theta)^{n}. \ \ (n>0)
	\end{equation}
	After a steady state is realized at $V_p$, we change the velocity to $10V_p$. This switch is done instantaneously.
	We also inspect the opposite case: switch from $10V_p$ to $V_p$.
	Such instantaneous switching processes are the ideal limit of infinite stiffness in Eqs. (\ref{power1}) and (\ref{power2}).

\begin{figure}
		\includegraphics[width=\linewidth]{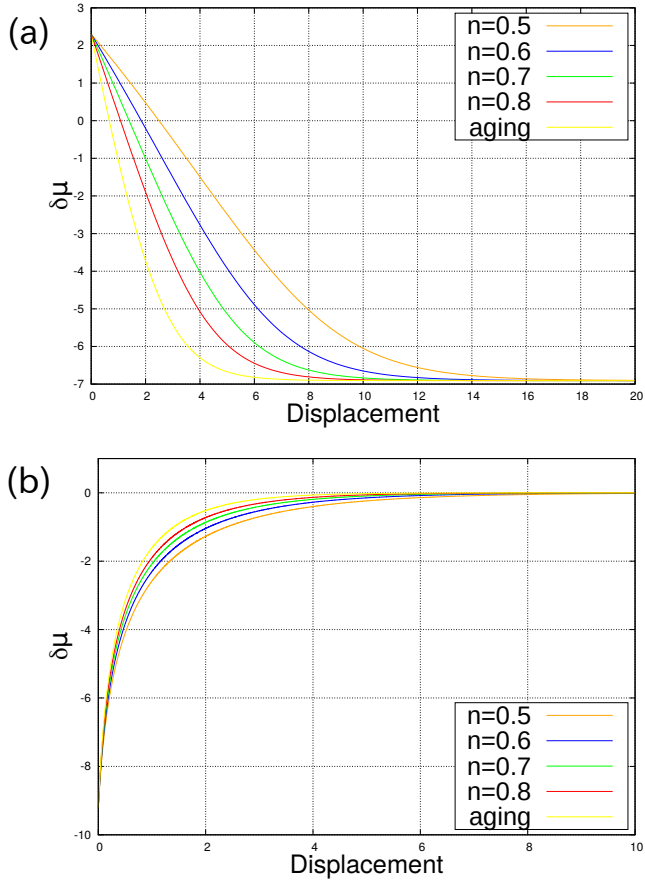}
		\caption{Relaxation of friction coefficient after an abrupt change of the slip velocity for modified ageing law I, calculated from Eqs. (\ref{deltamu}) and (\ref{nondimpow}) as a function of the (dimensionless) displacement. Parameters are set to $b=4.0$ and $a=1.0$. Effect of the exponent $n$ is studied by setting $n=0.5$, $0.6$, $0.7$, $0.8$, and $1.0$. 
		(a) Velocity is changed from $V=1.0$ to $V=10$. (b) Velocity is changed from $V=10$ to $V=1.0$.}
		\label{Relaxation_A}
	\end{figure}

	As shown in Figure \ref{Relaxation_A}, the relaxation after deceleration is generally steeper than that after acceleration. This asymmetric relaxation behaviour is common to the original ageing law, but the asymmetry is rather enhanced as the exponent $n$ decreases from $n=1$. In this respect, modified ageing law I is worse than the original one, although it produces the stable limit cycle.
 
   We also perform a slide-hold-slide test (Figures not shown), in which the static friction is measured as a function of the hold time. For $n>1$, this modified aging law predicts the static friction coefficient smaller than the experimental value. This may be expected from the PRZ law case \citep{Perrin1995, Marone1998}.
   For $n<1$, however, the predicted static friction coefficient is only slightly larger than that of the aging law, and therefore the modified aging law predicts the behaviour close to the experimental one.
   
	\subsection{Modified ageing law II}
\begin{figure}
		\includegraphics[width=\linewidth]{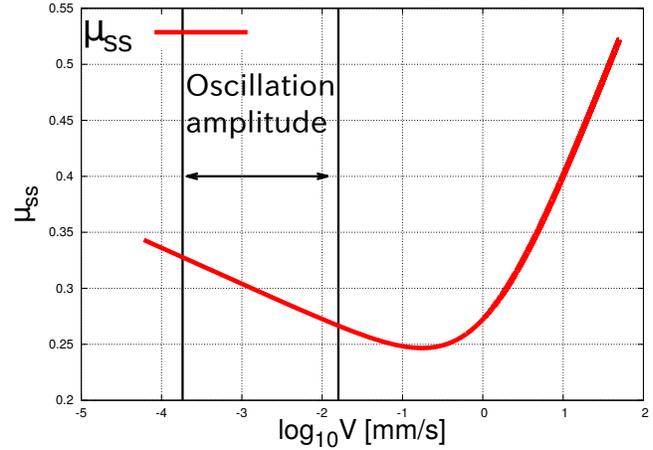}
		\caption{Velocity dependence of steady-state friction coefficient as calculated from the modified ageing law II, Eq. (\ref{SteadystateC}). The parameters are set to be $a=0.080$, $b=0.094$, $V_{c}=1.0\mathrm{[mm/s]}$, $c=0.5$, $\mu_{0}=0.369$ and $V_{0}=V_{p}=1.0\mathrm{[\mu m/s]}$. To emphasize that the velocity strengthening branch does not involve, the oscillation amplitude for $k=1.38\times 10^{-2}$ is shown, corresponding to Figure \ref{Cycle_B}.}
		\label{Steadystate}
	\end{figure}

	As we have seen, it is now possible to avoid the divergence problem with a minor change of the aging law. However, a modified aging law represented by Eq. (\ref{nondimpow}) does not work very well for velocity step experiments. More careful modification is thus needed.
	Here we propose another evolution equation.
	\begin{equation}
		\label{C}
		\dv{\hat{\theta}}{\hat{t}}=c+(1-c)\frac{\hat{V}}{V_{c}}-\frac{\hat{V}\hat{\theta}}{L}.
	\end{equation}
	Here $V_c$ is a characteristic velocity and $c$ is a constant satisfying $0<c<1$. At steady states, this equation leads to
	\begin{equation}
	    \theta_{\rm ss} = \frac{cL}{V}+(1-c)\frac{L}{V_c},
	\end{equation}
	and therefore
	\begin{equation}
		\label{SteadystateC}
		\mu_{\rm ss} = \mu_{0} + (a-b)\log(\frac{\hat{V}}{V_{0}}) + b\log\left[c+(1-c)\frac{\hat{V}}{V_{c}}\right].
	\end{equation}
	This velocity dependence, if $b>a$, is not monotonic as shown in Figure \ref{Steadystate}.
	At low velocities, the friction coefficient decreases in a logarithmic manner.
	This behaviour then changes at a specific velocity 
	\begin{equation}
	\label{crossoverV}
	\hat{V}=\frac{c}{1-c}\frac{b-a}{a} V_c,
	\end{equation}
	above which the velocity dependence is positive.
	A similar crossover is also observed in some experiments \citep{Heslot1994,Kilgore1993}. 
	The evolution law proposed here thus somehow captures this behaviour.
	
	Note that the parameters in this friction law can be estimated from experiments. Parameters $a$, $b$, and $L$ are estimated in a similar manner with the case of conventional evolution laws. Parameter $c$ should not be close to zero for healing to occur, while the possible range is $0<c<1$. In view of the experimental results by \cite{Heslot1994, Kilgore1993}, the constant $V_c$ may be on the order of $1$ to $10$ $\mathrm{[\mu m/s]}$. More detailed discussions are given in Appendix B.
	
	Applying this law to the single degree of freedom model with dimensionless variables defined by Eq. (\ref{nondim}), we get the following equations.
	\begin{equation}
		\label{C1}
		\dv{V}{t} = \frac{k}{a}(1-V)V-\frac{b}{a\theta}[c+(1-c)\alpha V-V\theta]V,
	\end{equation}
	\begin{equation}
		\label{C2}
		\dv{\theta}{t} = c+(1-c)\alpha V-V\theta.
	\end{equation}
	Here $\alpha \equiv V_{p}/V_{c}$. This system has the fixed point at $(V, \theta) = (1, c+(1-c)\alpha)$. At the point, Jacobi Matrix is
	\begin{equation}
		J = \left(
		\begin{array}{cc}
			\displaystyle{\frac{-k}{a}+\frac{bc}{a(c+(1-c)\alpha)}} &\displaystyle{\frac{b}{a}\frac{1}{c+(1-c)\alpha}}\\
			-c & -1
		\end{array}
		\right)
	\end{equation}
	Calculating the eigenvalues, we get the critical stiffness $k_c$ at which the Hopf bifurcation occurs.
	\begin{equation}
	\label{kcII}
	k_{c} = \frac{(b-a)c-a\alpha(1-c)}{c+(1-c)\alpha}.
	\end{equation}
	The first Lyapunov coefficient of this system is calculated as
	\begin{equation}
		l_{1}(0)=\frac{(c-1)\alpha b}{2\sqrt{a}[c+\alpha(1-c)]^3}\frac{1}{\sqrt{k_{c}}}.
	\end{equation}
	This is always negative since $0<c<1$, and therefore the system undergoes supercritical Hopf bifurcation.
	
\begin{figure}
		\includegraphics[width=\linewidth]{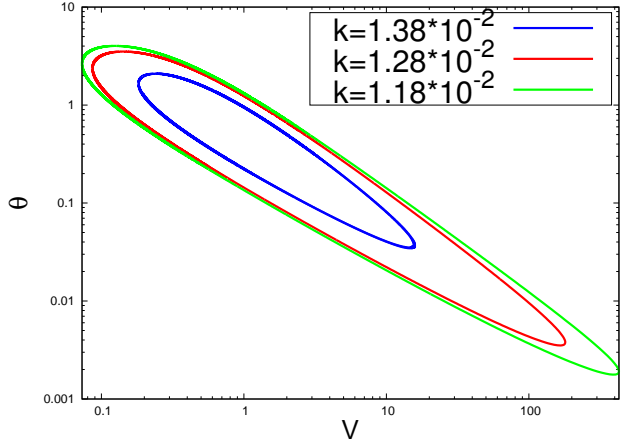}
		\caption{Limit cycles realized by Eqs. (\ref{C1}) and (\ref{C2}). Parameters are set to $a=0.08$, $b=0.094$, $c=0.5$, and $\alpha=0.001$, leading to $k_{c}=2.33$. Calculations are conducted for $k=1.38\times10^{-2}$, $1.28\times10^{-2}$ and $1.18\times10^{-2}$.}
		\label{Cycle_B}
	\end{figure}

	The existence of limit cycle is confirmed numerically as shown in Figure \ref{Cycle_B}, where the amplitude of the limit cycle increases as $k$ decreases. This is a clear evidence for supercritical Hopf bifurcation.
	We again wish to stress that radiation damping is not introduced here to simulate the limit cycle behaviour.
	
\begin{figure}
		\includegraphics[width=\linewidth]{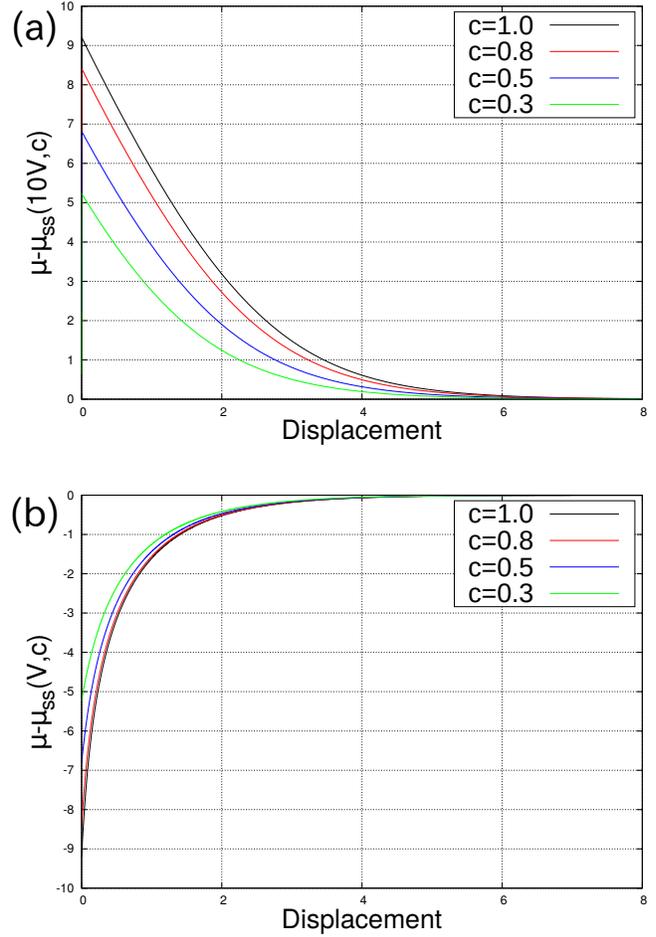}
		\caption{Relaxation of friction coefficient after an abrupt change of the slip velocity for modified ageing law II: Eqs. (\ref{deltamu}) and (\ref{C2}) with $c=0.8$, $0.5$, and $0.3$. Other parameters are set to be $a=1.0$, $b=4.0$, and $\alpha=0.1$. The vertical axis is set to be zero at the steady state after relaxation.
		The velocity is changed (a) from $V=1.0$ to $V=10$, (b) from $V=10$ to $V=1.0$. }
		\label{Relaxation_B}
	\end{figure}

	We then inspect the response to step-wise change of the slip velocity between $V_p$ and $10V_p$. Here we use nondimensional variables with (\ref{C2}). Results are shown in Figure \ref{Relaxation_B}. To clarify the relaxation behavior, the vertical axis is based on the steady state after relaxation. Although relaxation is not completely symmetric in terms of acceleration and deceleration, the deviation declines at smaller $c$.
	In short, this evolution law can realize almost the same degree of symmetry as the aging law. Although the symmetry is inferior to the slip law, there is enough rationality in using this law as a substitute for the aging law, considering that there was no evolution law that can realize the limit cycle.
	
	Note also, however, that this result is highly dependent on the parameters value. As can be seen from equation (\ref{C2}), if $\alpha$ is sufficiently small, the form of the equation is almost identical to the aging law, and the similar  behavior may appear regardless of the value of c.
	
	\section{Discussions}
	\subsection{Physical meaning of new evolution law}
	The evolution law given by Eq. (\ref{C}) may represent the effect of wear.
	Friction between brittle materials produces fine wear particles on the sliding interface.
	Wear particles have the fresh surfaces that possess adhesion force, changing the friction force significantly.
	However, the adhesion force may be lost due to oxidization of the surface.

	Here we assume that the surface of particles can be oxidized to loose the ability for healing, and incorporate this effect into the ageing law.
	The fraction of the oxidized surface is denoted by $1-c$, where $c$ is the fraction of fresh and unoxidized surface (thus $0<c<1$).
	The evolution law for the interface is at rest ($V=0$) then reads
	\begin{equation}
		\dv{\hat{\theta}}{\hat{t}} = c.
	\end{equation}
	This is interpreted that the ageing is suppressed due to oxidization of the surface.
	
	For sliding surfaces ($V>0$), wear particles are generated with the slip amount.
	These fresh particles have the unoxidized surface, and thus have the ability for healing.
	We assume that the oxidized surface is replaced by the fresh surface at the rate of $\hat V/V_c$, where $V_{c}$ is the characteristic velocity for wear.
	We thus have the additional ageing term $(1-c)\hat{V}/V_{c}$. 
	Combining these effects with the linear slip weakening term, we obtain Eq. (\ref{C}).
	
	\subsection{Oscillatory motion and velocity dependence of friction}
	With conventional evolution equations, radiation damping is an essential ingredient to reproduce the periodic oscillation in friction.
	Since it is similar to linear velocity-strengthening friction, one might regard the velocity-strengthening nature is essential  \citep{Matsuzawa2010, Shibazaki2007, Barbot2019}.
	However, we wish to stress that the velocity-strengthening friction is not essential to reproduce the periodic oscillation.
	This is obvious because Eq. (\ref{power}), which is velocity-weakening, reproduces the limit cycle.
	
	Similarly, another evolution equation, Eq. (\ref{C}), changes its velocity dependence from negative to positive (from velocity weakening to velocity-strengthening) at the crossover velocity given by Eq. (\ref{crossoverV}).
	However, we wish to stress that the limit cycle behaviour is not due to this crossover, because the velocity oscillates around $V_p$ and does not necessarily exceed the crossover velocity. In other words, the friction is always velocity-weakening during the oscillatory motion. The oscillatory motion is thus not due to the non-monotonic nature of the velocity dependence of friction.
	To illustrate this point more clearly, the velocity range of the limit cycle oscillation is shown in Figure \ref{Steadystate}.
	
	\subsection{Relevance to slow slip events}
	Recent development of observation technology enables us to discover various kinds of slow earthquakes \citep{Obara2016}, which span a wide range of time and energy scales \citep{Ide2007}. This implies that the natural fault motion is much more complex than a simple stick-slip motion. A remarkable case is off coast of Tohoku, Japan, where the subduction rate oscillates with the period of several years \citep{Uchida2016}. The maximum subduction rate during an oscillation period is only several times larger than the average subduction rate, and therefore this oscillation may be interpreted as a family of slow slip events (SSE).

	Here we wish to discuss SSE based on the present fault-patch model with the RSF. This simple model may be partly verified by observations that SSEs repeat themselves in patch-like regions. The use of RSF may be rational in view of small slip rates of SSE. Here we estimate the oscillation period using our new evolution equation, Eq. (\ref{C}).
	Linear stability analysis leads to the expression for the oscillation period, $T_2$.
	\begin{equation}
		T_{2}= 2\pi \frac{L}{V_{p}} \sqrt{\frac{a(c+(1-c)\alpha)}{(b-a)c-a\alpha(1-c)}},
	\end{equation}
	where $\alpha=V_{p}/V_{c}$.
	In a similar manner, the oscillation period for the aging law and the slip law is calculated as
	\begin{equation}
		T_{1}= 2\pi \frac{L}{V_{p}} \sqrt{\frac{a}{b-a}}.
	\end{equation}
	The ratio is
	\begin{equation}
		\frac{T_{2}}{T_{1}} = \sqrt{1+\frac{b\alpha (1-c)}{(b-a)c-a\alpha (1-c)}}.
	\end{equation}
	Recalling the critical spring constant $k_c$ given by Eq. (\ref{kcII}), $(b-a)c > a\alpha (1-c)$ must hold since $k_c$ must be positive. It thus follows that $T_{2}/T_{1} > 1$, meaning that a new evolution equation gives a longer period than the slip law or the ageing law.
	
	To estimate the period $T_1$ and $T_2$ for subduction zones, numerical values are required for $a$, $b$, $V_{p}$, and $L$. Precise estimate of these parameters for natural faults are difficult and there has been no agreement to date. We thus limit ourselves to a crude guess.
	The length constant $L$ has been estimated as several millimeters via the geodetic inversion on the afterslip data for the 2003 Tokachi-oki earthquake \citep{Fukuda2009}.
	We thus assume $L$ to be a millimeter here.
	The subduction velocity $V_{p}$ may be of the order of $\mathrm{[cm/year]}$.
	We assume it as several nanometers per second.
	Then $L/V_p$ is on the order of $10^6$ s.
	If we assume that $a$ and $b$ are on the order of $10^{-2}$ in view of typical experiments, $T_2$ is estimated as a few years, whereas $T_1$ is several months. In this respect, $T_2$ is more appropriate for the period of SSE cycles. Namely, the new evolution equation proposed here may be more suitable in explaining a period of several years for SSE.
	The period $T_1$ can be a few years if $b/a - 1$ is as small as $10^{-3}$, but such a condition requires finer tuning of the parameters $a$ and $b$.
	

    At the same time, however, we need to recall that the oscillation period estimated here is based on a simple fault-patch model. Subduction zones, where most slow earthquakes occur, extend over several hundred kilometers and involve many kinds of heterogeneity. If there is a slip patch that is sufficiently isolated from other seismogenic zones, the discussion here may apply. Otherwise, the discussion here should be interpreted as an ideal case analysis.
    
    Since the observed oscillation of subduction rates is spatially heterogeneous \citep{Uchida2016}, the discussion given here should be interpreted as a preliminary guess for more quantitative studies on spatially extended systems, which would be a promising attempts in the future.
    	
    \subsection{Case of multiplicative form}
    The discussions so far are based on the standard form of RSF, in which the effects of slip rate and state variable(s) are expressed in the additive manner.
    On the other hand, an alternative form is also known, in which the effects of slip rate and state variable(s) are expressed in the multiplicative manner.
    We may discuss the stability of limit cycle produced by this alternative form of RSF.
    
    We inspect two variations proposed by \cite{Barbot2019b} and \cite{Bar-Sinai2014}, respectively, by performing numerical calculations on the patch model together with the standard ageing law. We find the stable limit cycle without radiation damping for both the cases. Details of the results are given in the supplementary material. This result indicates the supremacy of the multiplicative form for the RSF law in reproducing the stable limit cycle. One may thus need to reflect on the use of popular additive form of the RSF, depending on the specific situation that one needs to consider.

	\section{Conclusions}
	In a slip patch model implemented with the rate and state friction (RSF) law, it has been well known that the numerical solution diverges unless the radiation damping is introduced.
	However, this divergence has been known only empirically and numerically.
	Here we have analytically shown that the divergence is inevitable for three major evolution equations: the slip law, the ageing law, and the Nagata's law.
	
	The radiation damping, which has been conventionally used, is due to the energy dissipation caused by the radiation of elastic waves. Therefore, from a physical point of view, introducing radiation damping may not be verified in models for slow aseismic motion such as slow slip events, in which seismic waves are not radiated. On the other hand, slow slip events are often accompanied by tremors, which radiate weak seismic waves. Such a complex nature of slow earthquakes may verify the use of radiation damping, but the quantitative aspect is not clear.
	
	In light of the above points, the choice of evolution equation requires discretion in modelling slow earthquakes, which radiate only a little amount of seismic waves. To model slow aseismic slip motion, two new evolution laws are proposed here. Both the equations reproduce slow periodic oscillation (i.e., the limit cycle) without radiation damping, at least for a single degree-of-freedom system. Modified ageing law I illustrates that a limit cycle can be realized only with a minor change in the evolution law. However, in view of the asymmetric relaxation after the abrupt velocity change, it cannot be a good replacement for the existing evolution laws. Modified ageing law II also exhibits asymmetric relaxation, but it is at the same level as the original ageing law. The only defect may be time-dependent healing is somewhat (by factor $c$) smaller than the original ageing law, but the difference is as small as $b \log c$. This is generally negligible.
	
	The difference between the two novel equations is most apparent in the steady-state behaviours. The velocity dependence of modified ageing law I is monotonic, whereas the modified ageing law II exhibits crossover to velocity-strengthening above a certain slip velocity. The difference may be useful in determining a suitable evolution law based on the experimental data. Regarding a relaxation behavior after an abrupt change of the slip velocity, the modified ageing law II is more consistent with experiments than the modified ageing law I.
	
	
	Since this study is limited to a single degree of freedom system, application of these new evolution laws to spatially extended systems may be promising future work toward understanding of slow earthquakes. At the same time, critical consideration on the use of additive form of the RSF is needed.

	\begin{acknowledgments}
	This study was supported by Japan Society for the Promotion of Science (JSPS) Grants-in-Aid for Scientific Research (KAKENHI) Grants Nos. JP16H06478 and 19H01811. Additional support from the Ministry of Education, Culture, Sports, Science and Technology (MEXT) of Japan, under its Earthquake and Volcano Hazards Observation and Research Program, is also gratefully acknowledged.
	\end{acknowledgments}
	
	\section*{Data Availability Statements}
	Simulation codes are available upon request.

	\bibliographystyle{gji}
	\bibliography{bib.bib}

\begin{thebibliography}{31}
\expandafter\ifx\csname natexlab\endcsname\relax\def\natexlab#1{#1}\fi

\bibitem[Bar-Sinai et~al.(2014)Bar-Sinai, Spatschek, Brener, \&
  Bouchbinder]{Bar-Sinai2014}
Bar-Sinai, Y., Spatschek, R., Brener, E.~A., \& Bouchbinder, E., 2014.
\newblock On the velocity-strengthening behavior of dry friction, {\it Journal
  of Geophysical Research: Solid Earth\/}, {\bf 119}(3), 1738--1748.

\bibitem[Barbot(2019{\natexlab{a}})]{Barbot2019}
Barbot, S., 2019{\natexlab{a}}.
\newblock Slow-slip, slow earthquakes, period-two cycles, full and partial
  ruptures, and deterministic chaos in a single asperity fault, {\it
  Tectonophysics\/}, {\bf 768}, 228171.

\bibitem[Barbot(2019{\natexlab{b}})]{Barbot2019b}
Barbot, S., 2019{\natexlab{b}}.
\newblock Modulation of fault strength during the seismic cycle by grain-size
  evolution around contact junctions, {\it Tectonophysics\/}, {\bf 765},
  129--145.

\bibitem[Beeler et~al.(1994)Beeler, Tullis, \& Weeks]{Beeler1994}
Beeler, N.~M., Tullis, T.~E., \& Weeks, J.~D., 1994.
\newblock The roles of time and displacement in the evolution effect in rock
  friction, {\it Geophysical Research Letters\/}, {\bf 21}(18), 1987--1990.

\bibitem[Bizzarri(2011)]{Bizzarri2011}
Bizzarri, A., 2011.
\newblock On the deterministic description of earthquakes, {\it Reviews of
  Geophysics\/}, {\bf 49}(3).

\bibitem[Cochard \& Madariaga(1994)]{Cochard1994}
Cochard, A. \& Madariaga, R., 1994.
\newblock Dynamic faulting under rate-dependent friction, {\it pure and applied
  geophysics\/}, {\bf 142}(3), 419--445.

\bibitem[Dieterich(1979)]{Dietrich1979}
Dieterich, J.~H., 1979.
\newblock Modeling of rock friction: 1. experimental results and constitutive
  equations, {\it Journal of Geophysical Research: Solid Earth\/}, {\bf
  84}(B5), 2161--2168.

\bibitem[Fukuda et~al.(2009)Fukuda, Johnson, Larson, \& Miyazaki]{Fukuda2009}
Fukuda, J., Johnson, K.~M., Larson, K.~M., \& Miyazaki, S., 2009.
\newblock Fault friction parameters inferred from the early stages of afterslip
  following the 2003 tokachi-oki earthquake, {\it Journal of Geophysical
  Research: Solid Earth\/}, {\bf 114}(B4).

\bibitem[Gu et~al.(1984)Gu, Rice, Ruina, \& Tse]{Gu1984}
Gu, J.-C., Rice, J.~R., Ruina, A.~L., \& Tse, S.~T., 1984.
\newblock Slip motion and stability of a single degree of freedom elastic
  system with rate and state dependent friction, {\it Journal of the Mechanics
  and Physics of Solids\/}, {\bf 32}(3), 167--196.

\bibitem[Hawthorne \& Rubin(2013)]{Hawthorne2013}
Hawthorne, J.~C. \& Rubin, A.~M., 2013.
\newblock Laterally propagating slow slip events in a rate and state friction
  model with a velocity-weakening to velocity-strengthening transition, {\it
  Journal of Geophysical Research: Solid Earth\/}, {\bf 118}(7), 3785--3808.

\bibitem[Heslot et~al.(1994)Heslot, Baumberger, Perrin, Caroli, \&
  Caroli]{Heslot1994}
Heslot, F., Baumberger, T., Perrin, B., Caroli, B., \& Caroli, C., 1994.
\newblock Creep, stick-slip, and dry-friction dynamics: Experiments and a
  heuristic model, {\it Phys. Rev. E\/}, {\bf 49}, 4973--4988.

\bibitem[Ide et~al.(2007)Ide, Beroza, Shelly, \& Uchide]{Ide2007}
Ide, S., Beroza, G.~C., Shelly, D.~R., \& Uchide, T., 2007.
\newblock A scaling law for slow earthquakes, {\it Nature\/}, {\bf 447}(7140),
  76--79.

\bibitem[Kilgore et~al.(1993)Kilgore, Blanpied, \& Dieterich]{Kilgore1993}
Kilgore, B.~D., Blanpied, M.~L., \& Dieterich, J.~H., 1993.
\newblock Velocity dependent friction of granite over a wide range of
  conditions, {\it Geophysical Research Letters\/}, {\bf 20}(10), 903--906.

\bibitem[Kuznetsov(2004)]{Kuznetsov2004}
Kuznetsov, Y.~A., 2004.
\newblock {\it Elements of applied bifurcation theory\/}, Springer-Verlag New
  York.

\bibitem[Marone(1998)]{Marone1998}
Marone, C., 1998.
\newblock Laboratory-derived friction laws and their application to seismic
  faulting, {\it Annual Review of Earth and Planetary Sciences\/}, {\bf 26}(1),
  643--696.

\bibitem[Matsuzawa et~al.(2010)Matsuzawa, Hirose, Shibazaki, \&
  Obara]{Matsuzawa2010}
Matsuzawa, T., Hirose, H., Shibazaki, B., \& Obara, K., 2010.
\newblock Modeling short- and long-term slow slip events in the seismic cycles
  of large subduction earthquakes, {\it Journal of Geophysical Research: Solid
  Earth\/}, {\bf 115}(B12).

\bibitem[Nagata et~al.(2012)Nagata, Nakatani, \& Yoshida]{Nagata2012}
Nagata, K., Nakatani, M., \& Yoshida, S., 2012.
\newblock A revised rate- and state-dependent friction law obtained by
  constraining constitutive and evolution laws separately with laboratory data,
  {\it Journal of Geophysical Research: Solid Earth\/}, {\bf 117}(B2).

\bibitem[Obara \& Kato(2016)]{Obara2016}
Obara, K. \& Kato, A., 2016.
\newblock Connecting slow earthquakes to huge earthquakes, {\it Science\/},
  {\bf 353}(6296), 253--257.

\bibitem[Perrin et~al.(1995)Perrin, Rice, \& Zheng]{Perrin1995}
Perrin, G., Rice, J.~R., \& Zheng, G., 1995.
\newblock Self-healing slip pulse on a frictional surface, {\it Journal of the
  Mechanics and Physics of Solids\/}, {\bf 43}(9), 1461--1495.

\bibitem[Ranjith \& Rice(1999)]{Ranjith1999}
Ranjith, K. \& Rice, J., 1999.
\newblock Stability of quasi-static slip in a single degree of freedom elastic
  system with rate and state dependent friction, {\it Journal of the Mechanics
  and Physics of Solids\/}, {\bf 47}(6), 1207--1218.

\bibitem[Rice \& Ruina(1983)]{RiceRuina1983}
Rice, J. \& Ruina, A., 1983.
\newblock Stability of steady frictional slipping, {\it Journal of Applied
  Mechanics-transactions of The Asme - J APPL MECH\/}, {\bf 50}, 343--349.

\bibitem[Rice(1993)]{Rice1993}
Rice, J.~R., 1993.
\newblock Spatio-temporal complexity of slip on a fault, {\it Journal of
  Geophysical Research: Solid Earth\/}, {\bf 98}(B6), 9885--9907.

\bibitem[Rice et~al.(2001)Rice, Lapusta, \& Ranjith]{Rice2001}
Rice, J.~R., Lapusta, N., \& Ranjith, K., 2001.
\newblock Rate and state dependent friction and the stability of sliding
  between elastically deformable solids, {\it Journal of the Mechanics and
  Physics of Solids\/}, {\bf 49}(9), 1865--1898, The JW Hutchinson and JR Rice
  60th Anniversary Issue.

\bibitem[Ruina(1983)]{Ruina1983}
Ruina, A., 1983.
\newblock Slip instability and state variable friction laws, {\it Journal of
  Geophysical Research: Solid Earth\/}, {\bf 88}(B12), 10359--10370.

\bibitem[Scholz(1990)]{Scholz}
Scholz, C.~H., 1990.
\newblock {\it The mechanics of earthquakes and faulting\/}, Cambridge
  university press.

\bibitem[Shibazaki \& Shimamoto(2007)]{Shibazaki2007}
Shibazaki, B. \& Shimamoto, T., 2007.
\newblock {Modelling of short-interval silent slip events in deeper subduction
  interfaces considering the frictional properties at the unstable?stable
  transition regime}, {\it Geophysical Journal International\/}, {\bf 171}(1),
  191--205.

\bibitem[Strogatz(2001)]{Strogatz2001}
Strogatz, S.~H., 2001.
\newblock {\it Nonlinear Dynamics And Chaos: With Applications To Physics,
  Biology, Chemistry, And Engineering (Studies in Nonlinearity)\/}, CRC Press.

\bibitem[Sugiura et~al.(2014)Sugiura, Hori, \& Kawamura]{Sugiura2012}
Sugiura, N., Hori, T., \& Kawamura, Y., 2014.
\newblock Synchronization of coupled stick-slip oscillators, {\it Nonlinear
  Processes in Geophysics\/}, {\bf 21}(1), 251--267.

\bibitem[Uchida et~al.(2016)Uchida, Iinuma, Nadeau, B{\"u}rgmann, \&
  Hino]{Uchida2016}
Uchida, N., Iinuma, T., Nadeau, R.~M., B{\"u}rgmann, R., \& Hino, R., 2016.
\newblock Periodic slow slip triggers megathrust zone earthquakes in
  northeastern japan, {\it Science\/}, {\bf 351}(6272), 488--492.

\bibitem[Viesca(2016{\natexlab{a}})]{Viesca2016}
Viesca, R.~C., 2016{\natexlab{a}}.
\newblock Self-similar slip instability on interfaces with rate- and
  state-dependent friction, {\it Proceedings of the Royal Society A:
  Mathematical, Physical and Engineering Sciences\/}, {\bf 472}(2192),
  20160254.

\bibitem[Viesca(2016{\natexlab{b}})]{Viesca2016PRE}
Viesca, R.~C., 2016{\natexlab{b}}.
\newblock Stable and unstable development of an interfacial sliding
  instability, {\it Phys. Rev. E\/}, {\bf 93}, 060202.

\end{thebibliography}


\begin{thebibliography}{5}
\expandafter\ifx\csname natexlab\endcsname\relax\def\natexlab#1{#1}\fi

\bibitem[Bar-Sinai et~al.(2014)Bar-Sinai, Spatschek, Brener, \&
  Bouchbinder]{Bar-Sinai2014}
Bar-Sinai, Y., Spatschek, R., Brener, E.~A., \& Bouchbinder, E., 2014.
\newblock On the velocity-strengthening behavior of dry friction, {\it Journal
  of Geophysical Research: Solid Earth\/}, {\bf 119}(3), 1738--1748.

\bibitem[Barbot(2019)]{Barbot2019b}
Barbot, S., 2019.
\newblock Modulation of fault strength during the seismic cycle by grain-size
  evolution around contact junctions, {\it Tectonophysics\/}, {\bf 765},
  129--145.

\bibitem[Kuznetsov(2004)]{Kuznetsov2004}
Kuznetsov, Y.~A., 2004.
\newblock {\it Elements of applied bifurcation theory\/}, Springer-Verlag New
  York.

\bibitem[Marone(1998)]{Marone1998}
Marone, C., 1998.
\newblock Laboratory-derived friction laws and their application to seismic
  faulting, {\it Annual Review of Earth and Planetary Sciences\/}, {\bf 26}(1),
  643--696.

\bibitem[Strogatz(2001)]{Strogatz2001}
Strogatz, S.~H., 2001.
\newblock {\it Nonlinear Dynamics And Chaos: With Applications To Physics,
  Biology, Chemistry, And Engineering (Studies in Nonlinearity)\/}, CRC Press.

\end{thebibliography}
	
	\appendix

	\section{Normal form theory}
	To know the type of bifurcation, we must calculate the nonlinear effect directly. Normal form theory is an effective way for this purpose. Here we consider a 2D dynamical system that undergoes Hopf bifurcation. According to the theory, after appropriate coordinate transformation, one can rewrite the time evolution equation as follows.
	in the neighbourhood of the bifurcation point,
	\begin{equation}
		\dv{u}{t} = (\beta+i)u+\sigma_{1}u\abs{u}^{2}+\order{\abs{u^{4}}},
	\end{equation}
	where $u$ is a complex variable, $\beta$ is a constant, and $\sigma_1 = \pm 1$ corresponding to the the sign of the first Lyapunov coefficient, $l_{1}(0)$. The functional form of $l_{1}(0)$ is not shown here. One can refer to a reference \citep{Kuznetsov2004} or supplemental material.
	
	The type of Hopf bifurcation, either supercritical or subcritical, is determined by $\sigma_{1}$. When $\sigma_{1} = -1$, the system undergoes supercritical Hopf bifurcation. In this case, the limit cycle of small amplitude arises once the system loses its stability. If $\sigma_{1} = 1$, subcritical Hopf bifurcation occurs. Then the limit cycle appears with the large amplitude. In general, the solution diverges once the system loses its stability.
	
	There are some methods to calculate the value of $l_{1}(0)$ numerically. In this paper, we employ the method stated in \citep{Kuznetsov2004} using a software, Maple$^{TM}$. Detailed calculation is shown in supplement material.
	
	\section{Determining the parameters for modified aging law II}
   The parameters in modified aging law II can be determined from experimental data.
   Noting that $\mu_{0}$ and $V_{0}$ are arbitrary constants, the parameters to be determined are $a, b, c, V_{c}$ and $L$. 
   
   \begin{enumerate}
   	\item  Parameter $a$ \\
   	This is determined from the direct effect in the velocity step experiment, as is usually done for major evolution laws. For sudden velocity change, the state variable does not change and the change of friction coefficient $\delta \mu$ is written as
   	\begin{equation}
   		\delta \mu =  a \log(\frac{V_{2}}{V_{1}}),
   	\end{equation}
   	where the slip velocity is changed from $V_{1}$ to $V_{2}$.

   	\item  Parameter $b$ \\
   		 Parameter $b$ involves the healing \citep{Beeler1994, Marone1998}. If holding time is sufficiently long and the apparatus is sufficiently stiff, the velocity can be regarded to be zero during the holding time. Then the state variable evolves as
   	\begin{equation}
   		\dv{\hat{\theta}}{\hat{t}} = c.
   	\end{equation}
   	Then, as in the case of original aging law, the amount of the healing of the friction $\Delta \mu$ depends on the holding time as
   	\begin{equation}
   		\dv{\Delta \mu}{\log \hat{t}} \simeq b,
   	\end{equation}
   	from which parameter $b$ is determined.
   	
   	\item  Parameters $c$ and $V_{c}$ \\
   	These parameters are determined from the crossover velocity $\hat{V}_{\rm cross}$ and the value of the friction coefficient there:
   	\begin{equation}
   		\label{cross}
   		\hat{V}_{\rm cross}=\frac{c}{1-c}\frac{b-a}{a} V_{c},
   	\end{equation}
   	above which the velocity dependence switches from negative to positive. The steady-state friction coefficient at the crossover velocity is
   	\begin{equation}
   	\label{mussVcross}
   		\mu_{\rm ss} = \mu_{0} + (a-b)\log(\frac{{\hat V}_{\rm cross}}{V_{0}}) +b\log(\frac{b}{a}) + b\log c.
   	\end{equation}
   	Since ${\hat V}_{\rm cross}$ can be known from experiment, $c$ and $V_c$ are found using Eqs. (\ref{cross}) and (\ref{mussVcross}).

   	\item  Parameter $L$ \\
   	This is determined from the relaxation process in velocity step tests by comparing the model behaviour with the observed relaxation process.
   \end{enumerate}.	

	\label{lastpage}
	
\end{document}


\label{firstpage}
 \section{Slide-Hold-Slide test for modified aging law I}	
To investigate the healing effect of the modified aging law I, we conduct the slide-hold-slide test. We solve the following equations 
\begin{equation}
	\dv{\hat{V}}{\hat{t}}=\frac{k^{'}}{a}\hat{V}(V_{p}-\hat{V})-\frac{b}{a\hat{\theta}}\left[1-\left(\frac{\hat{V}\hat{\theta}}{L}\right)^{n}\right]\hat{V},
\end{equation}
\begin{equation}
	\label{power}
	\dv{\hat{\theta}}{\hat{t}}=1-\left(\frac{\hat{V}\hat{\theta}}{L}\right)^{n}.
\end{equation}
Here $k^{'}$ is defined as $k^{'}\equiv \hat{k}/\sigma$.

In this test, steady sliding state is realized at the slip velocity of $V_{p}$ for the duration of $100 [\mathrm{s}]$. Here we adopt $V_p = 3.0 \times 10^{-7} [\mathrm{m/s}]$. Then the slip velocity is changed to zero and keep this state for a certain time. We then change the slip velocity to $V_{p}$ again. Through this process, change in friction $\Delta \mu$ is measured.

\begin{figure}
	\begin{center}
		\includegraphics[width=9.5cm]{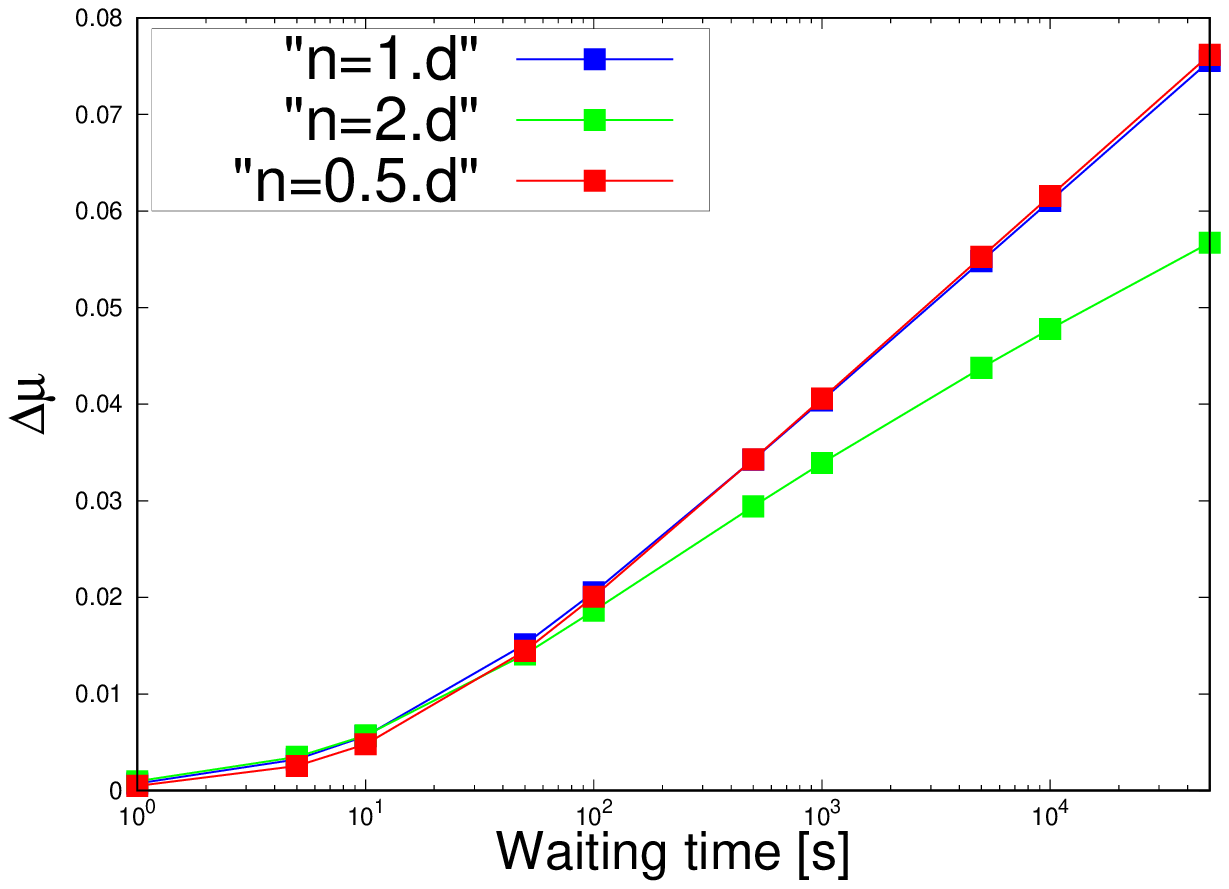}
		\caption{Results of the slide-hold-slide test for the modified aging law I. We show results for $n=0.5$, $1$, and $2$. Load point velocity is set to $V_{p}=3.0 \times 10^{-7} [\mathrm{m/s}]$. Initial sliding time is set to $100[\mathrm{s}]$. The other parameters are set to $a=8.0\times 10^{-3}$, $b=9.0\times 10^{-3}$, $L=3.0\times 10^{-6}{[\mathrm{m}]}$ and $k^{'}=7.4\times 10^{4}[\mathrm{m^{-1}}]$.}
		\label{waiting}
		\end{center}
	\end{figure}

We set parameters as $a=8.0\times 10^{-3}$, $b=9.0\times 10^{-3}$, $L=3.0\times 10^{-6}{[\mathrm{m}]}$ and $k^{'}=7.4\times 10^{4}[\mathrm{m^{-1}}]$. These parameter values are based on Fig.3 of \citep{Marone1998}. We test the case of $n=0.5$, $1$, and $2$.
The numerical results are shown in Fig.\ref{waiting}. Interpretation of these results is explained in the main text.
    
	\section{Dulac's criterion}
	For convenience, a brief proof of the Dulac's criterion is given based on \citep{Strogatz2001}.
	Consider a closed orbit $C$ and the area $A$ surrounded by $C$. We integrate $\nabla \vdot({g\vb{\dot{x}}})$ on $A$. 
	\begin{equation}
		\int_{A} \nabla \vdot({g\vb{\dot{x}}}) dxdy =\int_{A}\left[\pdv{(g\dot{x})}{x}+ \pdv{(g\dot{y})}{y}\right]dxdy.
	\end{equation}
	Using the Green's theorem, this is reduced to
	\begin{equation}
		=\oint_{C}(-g\dot{y} dx +g\dot{x} dy)=\oint_{C}(g\vb{\dot{x}}) \vdot d\vb{n},
	\end{equation}
	where $d\vb{n}=(dy, -dx)^{T}$ is orthogonal to the closed orbit $C$. 
	By definition, $\vb{\dot{x}}$ is tangent to $C$, and thus orthogonal to $d\vb{n}$.
	The above integral then vanishes, leading to 
	\begin{equation}
		\int_{A} \nabla \vdot({g\vb{\dot{x}}}) dxdy = 0.
	\end{equation} 
	This must hold for any area $A$ surrounded by a closed orbit.

	If the sign of the integrand $\nabla \vdot({g\vb{\dot{x}}})$ is always positive or negative on $A$, this equation cannot hold, and the boundary of $A$ cannot be a closed orbit. If this is true for any area, the closed orbit cannot exist on the entire space.
	
	\section{Normal form theory of Hopf bifurcation}
	In the main text, we have used a nonlinear analysis method called the normal form theory for Hopf bifurcation. Here we outline the theory. Our calculations are based on \citep{Kuznetsov2004}.
	
	First, consider the following two-dimensional differential equations that are on the critical point of Hopf bifurcation.
	\begin{equation}
		\dv{\xi_{1}}{t}=F_{1}(\xi_{1},\xi_{2}),
	\end{equation}
	\begin{equation}
		\dv{\xi_{2}}{t}=F_{2}(\xi_{1},\xi_{2}).
	\end{equation}
	For simplicity, we assume that the fixed point of this system is the origin. We then rewrite these equations in vector form as
	\begin{equation}
		\dv{\vb{\xi}}{t}=F(\vb{\xi}).
	\end{equation}
	Expanding the right hand side around the fixed point, one gets
	\begin{equation}
		\dv{\vb{\xi}}{t}=A\vb{\xi}+\frac{1}{2}B(\vb{\xi},\vb{\xi})+\frac{1}{6}C(\vb{\xi},\vb{\xi},\vb{\xi}),
	\end{equation}
	where $A$ is the Jacobi matrix at the fixed point, and $B$ and $C$ are the coefficients for the Taylor expansion. For instance, $B$ is given by
	\begin{equation}
		B_{l}(\vb{\xi},\vb{\eta})=\sum_{i,j}\left. \pdv[2]{F_{l}}{\xi_{i}}{\eta_{j}}\right|_{\xi=0,\eta=0}\xi_{i} \eta_{j}.
	\end{equation}
	and $C$ is given in a similar manner.
	
	The critical eingenvectors $\vb{q}$ and $\vb{p}$ are defined as
	\begin{equation}
		A\vb{q}=I\omega \vb{q},
	\end{equation}
	\begin{equation}
		A^{T}\vb{p}=-I\omega \vb{p}.
	\end{equation}
	The normalization condition is $\langle\vb{p},\vb{q}\rangle\equiv \sum_{i} \bar{p}_{i}q_{i} =1$, where $\bar{p_i}$ represents the complex conjugate of $p_i$. Using these vectors, we can write $\xi$ as
	\begin{equation}
		\label{coord}
		\xi=z\vb{q}+\bar{z}\bar{\vb{q}}.
	\end{equation}
	Substituting this in the first equation, one gets
	\begin{equation}
		\dv{(z\vb{q}+\bar{z}\bar{\vb{q}})}{t}=A*(z\vb{q}+\bar{z}\bar{\vb{q}})+\frac{1}{2}B(z\vb{q}+\bar{z}\bar{\vb{q}},z\vb{q}+\bar{z}\bar{\vb{q}})
		+\frac{1}{6}C(z\vb{q}+\bar{z}\bar{\vb{q}},z\vb{q}+\bar{z}\bar{\vb{q}},z\vb{q}+\bar{z}\bar{\vb{q}}).
	\end{equation}
	Taking the inner product with $p$, this equation becomes
	\begin{equation}
		\dv{z}{t}=I\omega z+\frac{1}{2}(\langle \vb{p},B(\vb{q},\vb{q})\rangle z^2+\langle \vb{p},B(\vb{q},\bar{\vb{q}})\rangle z\bar{z}+...)+\frac{1}{6}(\langle \vb{p},C(\vb{q},\vb{q},\vb{q})\rangle z^3+...).
	\end{equation}
	According to the normal form theory for Hopf bifurcation, by using a proper nonlinear transformation, this equation can be transformed into the following the normal form
	\begin{equation}
		\dv{u}{t} = (\beta+I)u+\sigma_{1}u\abs{u}^{2}+\order{\abs{u^{4}}},
	\end{equation}
	where $u$ is a complex variable, $\beta$ is a constant, and $\sigma_1 = \pm 1$ corresponds to the the sign of the first Lyapunov coefficient, $l_{1}(0)$. It is given by
	\begin{equation}
		\label{main}
		l_{1}(0)=\frac{1}{2\omega^2}\Re(Ig_{20}g_{11}+\omega g_{21}),
	\end{equation}
	where
	\begin{equation}
		g_{20}=\langle \vb{p},B(\vb{q},\vb{q})\rangle ,
	\end{equation}
	\begin{equation}
		g_{11}=\langle \vb{p},B(\vb{q},\bar{\vb{q}})\rangle ,
	\end{equation}
	\begin{equation}
		g_{21}=\langle \vb{p},C(\vb{q},\vb{q},\bar{\vb{q}})\rangle .
	\end{equation}
	
	Whether the Hopf bifurcation is supercritical or subcritical is determined by $\sigma_{1}$. If $\sigma_{1} = -1$, the system undergoes the supercritical Hopf bifurcation. In this case, the limit cycle of small amplitude arises once the system loses its stability. If $\sigma_{1} = 1$, the subcritical Hopf bifurcation occurs. Then the limit cycle appears with a large amplitude. The solution diverges once the system loses its stability.
	
	For simplicity, we assume that the fixed point is the origin. If the fixed point is not located at the origin, the fixed point must be displaced to the origin. For instance, if $\vb{\xi}^{0}$ is the fixed point instead of Eq.(\ref{coord}), one should substitute
	\begin{equation}
		\vb{\xi}=\vb{\xi}^{0}+z\vb{q}+\bar{z}\bar{\vb{q}}
	\end{equation}
	into the equations and expand the right-hand side around the fixed point. Then, as mentioned above, the nonlinear parts $B$ and $C$ are obtained.
	
	\section{Derivation of normal form for modified aging law I}
	One can calculate the first lyapunov coefficient for modified aging law I, which we derive as Eq.(32) in the main text.
	Governing equations are given as
	\begin{equation}
		\label{power1}
		\dv{V}{t}=\frac{k}{a}V(1-V)-\frac{b}{a\theta}\left[1-(V\theta)^{n}\right]V,
	\end{equation}
	\begin{equation}
		\label{power2}
		\dv{\theta}{t}=1-(V\theta)^{n}.
	\end{equation}
	Then let us calculate the Jacobi matrix $A$ and its transpose matrix $A^{T}$. Recalling that the fixed point is $(V,\theta)=(1,1)$ and that the critical point is $k=n(b-a)$, these matrices are derived as
	\begin{equation}
		A = \left[\begin{array}{cc}
			n & \frac{n^{2}+\omega^{2}}{n} 
			\\
			-n & -n 
		\end{array}\right],
	\end{equation}
	\begin{equation}
		A^{T}= \left[\begin{array}{cc}
			n & -n 
			\\
			\frac{n^{2}+\omega^{2}}{n} & -n 
		\end{array}\right].
	\end{equation}
	One can use the critical frequency $\omega$ defined as
	\begin{equation}
		\label{crit1}
		\omega^{2} = \frac{n^{2} \left(b-a\right)}{a}.
	\end{equation}
	From these conditions, one can get the critical eigenvectors as 
	\begin{equation}
		\vb{q} = \left[\begin{array}{c}
			\frac{n^{2}+\omega^{2}}{n \left(\mathrm{I} \omega-n\right)} 
			\\
			1 
		\end{array}\right],
	\end{equation}
	\begin{equation}
		\vb{p}=\left[\begin{array}{c}
			-\frac{\left(\mathrm{I} n-\omega\right) n}{2 \omega \left(\mathrm{I} \omega+n\right)} 
			\\
			\frac{-\mathrm{I} n+\omega}{2 \omega} 
		\end{array}\right].
	\end{equation}
	Here the coefficients of the vector $\vb{p}$ are defined to satisfy the conditions: $\langle \vb{p},\vb{q}\rangle =1$.
	
	Then let us calculate the nonlinear part $B$ and $C$. This calculation is somewhat cumbersome and therefore one may use any computer algebra system. (Here we use Maple.) Each term of Eq.(\ref{main}) are then given as
	\begin{equation}
		g_{2,0} = \frac{\left(-4 \,\mathrm{I} b+4 \,\mathrm{I} a\right) n^{3}+\omega \left(\left(\mathrm{I} a-\mathrm{I} b\right) \omega-4 a+2 b\right) n^{2}-\left(\mathrm{I} a-\mathrm{I} b-\omega a\right) \omega^{2} n-a \omega^{3}}{2 n \omega a},
	\end{equation}
	\begin{equation}
		g_{1,1}= \frac{\left(-4 \,\mathrm{I} b+4 \,\mathrm{I} a\right) n^{3}-\omega \left(\left(\mathrm{I} a-\mathrm{I} b\right) \omega-2 a\right) n^{2}+\left(-a \omega^{3}+\left(3 \,\mathrm{I} a-\mathrm{I} b\right) \omega^{2}\right) n+a \omega^{3}}{2 n \omega a},
	\end{equation}	
	\footnotesize
	\begin{equation}
		g_{2,1}= \frac{\mathrm{I} a \left(n-2\right) \left(n-1\right) \omega^{4}-\left(n \left(-b+a\right)-b\right) \left(n-1\right) n \omega^{3}-4 \,\mathrm{I} a \left(n-1\right) n^{2} \omega^{2}+2 \left(n \left(-b+a\right)-a\right) n^{3} \omega+12 \,\mathrm{I} n^{4} b}{2 n^{2} \omega a}.
	\end{equation}
	\normalsize
	Thus, the first Lyapunov coefficient is
	\begin{equation}
		l_{1}(0) = \frac{\left(\left(-b+a\right) n^{4}-\frac{\omega^{2} \left(-b+a\right) n^{3}}{4}+\frac{5 \omega^{2} \left(a-\frac{b}{5}\right) n^{2}}{4}-\frac{a n \omega^{4}}{2}+\frac{a \omega^{4}}{2}\right) \left(-b+a\right)}{\omega^{3} n \,a^{2}}.
	\end{equation}
	Substituting Eq.(\ref{crit1}) into the above equation, one gets
	\begin{equation}
		l_{1}(0) = \frac{1}{4}\left(\frac{b-a}{a}\right)^\frac{3}{2}(n-1),
	\end{equation}
	which is Eq.(32) of the main text.
	
	\section{Derivation of normal form for modified aging law II}
	One can calculate the first lyapunov coefficient for modified aging law II, which is Eq.(43) of the main text. 
	Governing equations are given as 
	\begin{equation}
		\label{C1}
		\dv{V}{t} = \frac{k}{a}(1-V)V-\frac{b}{a\theta}[c+(1-c)\alpha V-V\theta]V,
	\end{equation}
	\begin{equation}
		\label{C2}
		\dv{\theta}{t} = c+(1-c)\alpha V-V\theta.
	\end{equation}
	
	Let us calculate Jacobi matrix $A$ and its transpose matrix $A^{T}$. Recalling that the fixed point is $(V, \theta) = (1, c+(1-c)\alpha)$ and that the critical point is $k= ((b-a)c-a\alpha(1-c))/(c+(1-c)\alpha) \equiv k_{c}$, one can get
	\begin{equation}
		A = \left[\begin{array}{cc}
			1 & \frac{\omega^{2}+1}{c} 
			\\
			-c & -1 
		\end{array}\right],
	\end{equation}
	\begin{equation}
		A^{T} = \left[\begin{array}{cc}
			1 & -c 
			\\
			\frac{\omega^{2}+1}{c} & -1 
		\end{array}\right].
	\end{equation}
	Here we use the critical frequency $\omega$ defined as
	\begin{equation}
		\label{crit2}
		\omega^2	 = \frac{\left(-\alpha a+a-b\right) c+\alpha a}{\left(\left(\alpha-1\right) c-\alpha\right) a}.
	\end{equation}
	From these conditions, one can get critical eigenvectors as 
	\begin{equation}
		\vb{q} = \left[\begin{array}{c}
			\frac{\omega^{2}+1}{c \left(\mathrm{I} \omega-1\right)} 
			\\
			1 
		\end{array}\right],
	\end{equation}
	\begin{equation}
		\vb{p} = \left[\begin{array}{c}
			-\frac{\left(\mathrm{I}-\omega\right) c}{2 \omega \left(\mathrm{I} \omega+1\right)} 
			\\
			\frac{-\mathrm{I}+\omega}{2 \omega} 
		\end{array}\right].
	\end{equation}
	Here the coefficients of the vector $\vb{p}$ is defined to satisfy the normalization condition: $\langle \vb{p},\vb{q}\rangle =1$.
	
	Each term of Eq.(\ref{main}) is then given as
	\footnotesize
	\begin{equation}
		g_{2,0}=\frac{\left(2 a \left(\mathrm{I}-\omega\right) \alpha^{2}-4 \left(a-\frac{b}{2}\right) \left(\mathrm{I}-\omega\right) \alpha+\left(2 \,\mathrm{I}-2 \omega\right) a-2 \left(\mathrm{I}-\frac{\omega}{2}\right) b\right) c^{2}-4 \left(\alpha a-a+\frac{1}{2} b\right) \alpha \left(\mathrm{I}-\omega\right) c+2 a \left(\mathrm{I}-\omega\right) \alpha^{2}}{c a \left(\left(\alpha-1\right) c-\alpha\right)^{2} \omega},
	\end{equation}
	\normalsize
	\begin{equation}
		g_{1,1}= \frac{\left(\left(\alpha-1\right) c-\alpha\right) \left(\mathrm{I} \omega^{2}+2 \,\mathrm{I}+\omega\right) a+2 \,\mathrm{I} b c}{c a \left(\left(\alpha-1\right) c-\alpha\right) \omega},
	\end{equation}
	
	\begin{equation}
		g_{2,1} = -\frac{\left(\left(c-1\right)^{2} \left(-2 \omega+3 \,\mathrm{I}+\mathrm{I} \omega^{2}\right) \alpha^{2}-c \left(c-1\right) \left(\mathrm{I} \omega^{2}+9 \,\mathrm{I}-4 \omega\right) \alpha+6 c^{2} \left(\mathrm{I}-\frac{\omega}{6}\right)\right) b}{c a \left(\left(c-1\right) \alpha-c\right)^{3} \omega}.
	\end{equation}
	Substituting the above equations and Eq.(\ref{crit2}) into Eq.(\ref{crit1}), the first lyapunov coefficient is derived as
	\begin{equation}
		l_{1}(0) = -\frac{\left(c-1\right) \alpha b}{2 \mathrm{\sqrt{a}}\, \sqrt{\frac{-\left(c-1\right) a \alpha+c \left(a-b\right)}{\left(c-1\right) \alpha-c}}\, \left(\alpha c-\alpha-c\right)^{3}}
		=\frac{(c-1)\alpha b}{2\sqrt{a}[c+\alpha(1-c)]^3}\frac{1}{\sqrt{k_{c}}}.
	\end{equation}

 \section{Multiplicative RSF law}
 In the discussion part of the main text, the stability of limit cycle produced by the multiplicative RSF law is pointed out. Here we describe the details of calculation.
  \subsection{\cite{Barbot2019b}}
  	\cite{Barbot2019b} suggests the following equation.
  \begin{equation}
    \mu = \mu_{0}(\frac{\hat{V}}{V_{0}})^{\frac{a}{\mu_{0}}}(\frac{\hat{\theta} V_{0}}{L})^{\frac{b}{\mu_{0}}}.
  \end{equation}
  Applying this equation to the fault patch model with the aging law, the governing equations can be derived as
  \begin{equation}
  	\dv{V}{t}=\frac{k}{a V^\frac{a}{\mu_{0}}\theta^\frac{b}{\mu_{0}}}V(1-V)-\frac{b}{a\theta}(1-V\theta)V
  \end{equation}
  \begin{equation}
  	\dv{\theta}{t}=1-V\theta
  \end{equation}
Here we use non-dimensional variables defined in the main paper and set $V_{0}=V_{p}$. This system has a fixed point at $(V, \theta)=(1, 1)$. According to the linear stability analysis, the Hopf bifurcation point is $k_{c}=b-a$. Note also that the steady-state friction coefficient is
\begin{equation}
   \mu_{ss} \propto V^{\frac{a-b}{\mu_{0}}}.
\end{equation}

\begin{figure}
	\begin{center}
		\includegraphics[width=9.5cm]{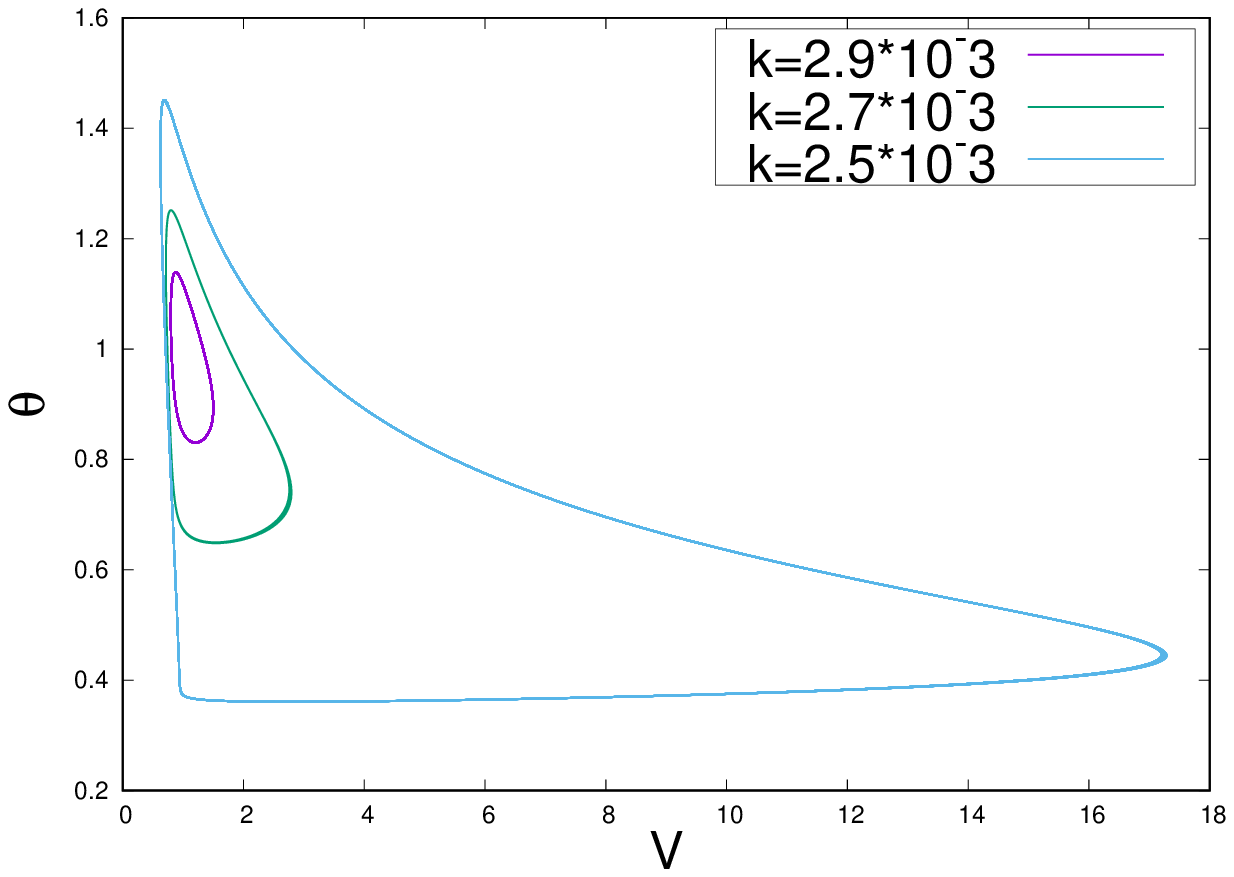}
		\caption{The limit cycle realized by the multiplicative RSF law proposed in \cite{Barbot2019b}. We show results for  $k=2.9\times10^{-3}, 2.7\times10^{-3}$ and $2.5\times10^{-3}$. The other parameters are set to $a=1.0\times 10^{-3}$, $b=4.0\times 10^{-3}$ and $\mu_{0}=1.0\times 10^{-3}$. The critical point is $k_{C}=3.0\times 10^{-3}$}
		\label{Cycle1}
	\end{center}
\end{figure}

We solve this model numerically. Parameters are set as $a=1.0\times10^{-3}$, $b=4.0\times10^{-3}$ and $\mu_{0}=1.0\times10^{-3}$, leading to $k_{c}=3.0\times10^{-3}$. Results for $k=2.9\times10^{-3}, 2.7\times10^{-3}$ and $2.5\times10^{-3}$ are shown in Fig.\ref{Cycle1}. In this case, we acknowledge the existence of the limit cycle.
 
 \subsection{\cite{Bar-Sinai2014}}
	\cite{Bar-Sinai2014} suggests the following equation.
	\begin{equation}
		\mu = f_{0}+\alpha\log(\frac{\hat{V}}{V_{0}})+\beta\log(1+\frac{\hat{\theta} V_{0}}{L})+\frac{\alpha \beta}{f_{0}}\log(\frac{\hat{V}}{V_{0}})\log(1+\frac{\hat{\theta} V_{0}}{L}).
	\end{equation}
 The equation is originally proposed for steady-state, but here we assume its general validity. Meaning of the parameters are stated in \citep{Bar-Sinai2014}. Applying this equation to the fault patch model with the aging law, governing equations are derived as
\begin{equation}
	\dv{V}{t}=	\frac{k}{\alpha +\frac{\alpha\beta}{f_{0}}\log\left(1+\theta\right)}(1-V)V
	-\frac{\left(1-V \theta \right) \left(\beta +\frac{\alpha\beta}{f_{0}} \log V \right)}{(\alpha +\frac{\alpha\beta}{f_{0}} \log \left(1+\theta \right))(1+\theta)}V
\end{equation}

\begin{equation}
	\dv{\theta}{t}=1-V\theta
\end{equation}
Here we use non-dimensional variables defined in the main paper and set $V_{0}=V_{p}$. This system has a fixed point at $(V, \theta)=(1, 1)$. According to the linear stability analysis, the Hopf bifurcation point is $k_{c}=\beta/2-\alpha-(\alpha\beta/f_{0})\log2$. 

\begin{figure}
	\begin{center}
		\includegraphics[width=9.5cm]{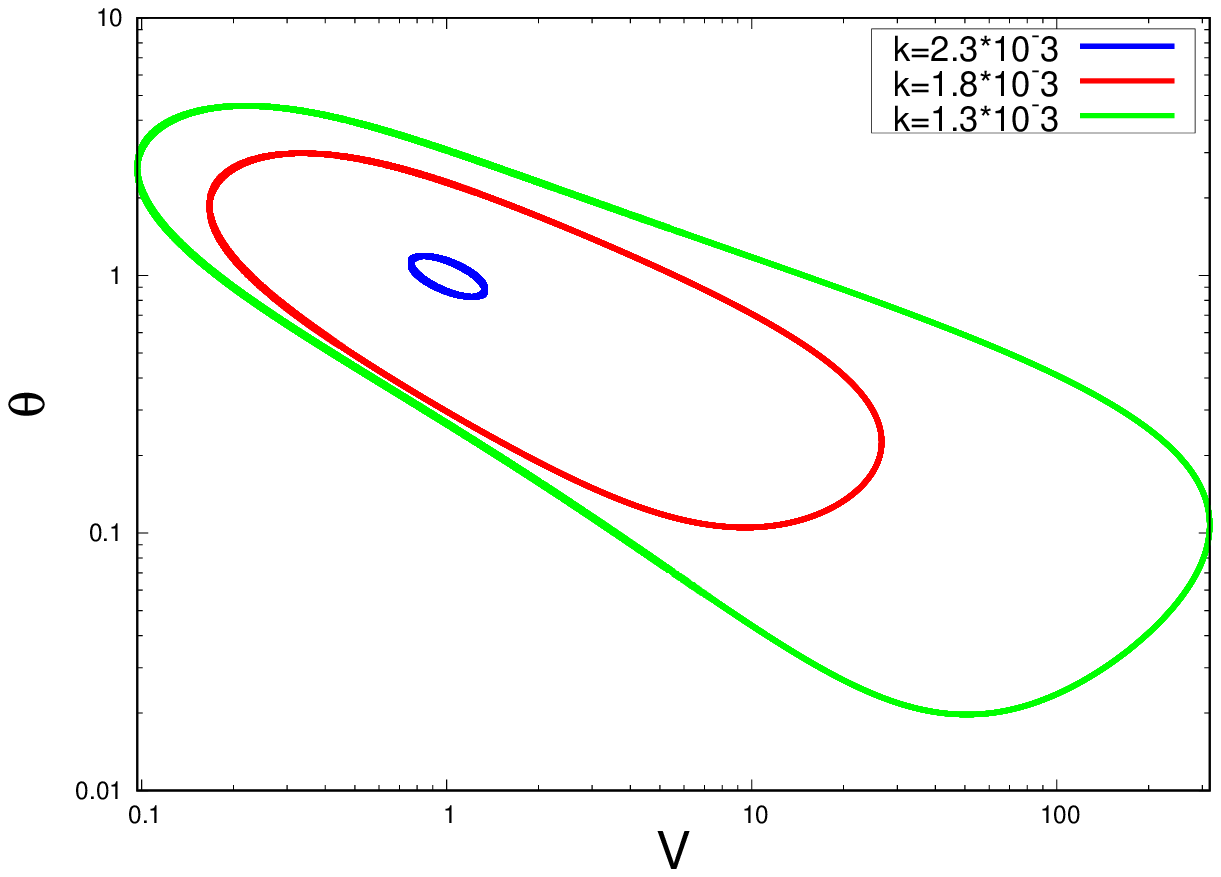}
		\caption{The limit cycle realized by  multiplicative RSF law proposed in \citep{Bar-Sinai2014}. We show results for $k=2.3\times10^{-3}, 1.8\times10^{-3}$ and $1.3\times10^{-3}$. The other parameters are set to $\alpha=1.0\times 10^{-3}$, $\beta=8.0\times 10^{-3}$ and $f_{0}=8.0\times 10^{-3}$. The critical point is $k_{C}=2.31\times 10^{-3}$}
		\label{Cycle2}
	\end{center}
\end{figure}

We solve the model numerically. Parameters are set as $\alpha=1.0\times10^{-3}$, $\beta=8.0\times10^{-3}$ and $f_{0}=8.0\times10^{-3}$, leading to $k_{c}=2.31\times10^{-3}$. Results for $k=2.3\times10^{-3}, 1.8\times10^{-3}$ and $1.3\times10^{-3}$ are shown in Fig.\ref{Cycle2}. We also acknowledge the existence of the limit cycle in this model.

	\label{lastpage}
	\bibliographystyle{gji}
	\bibliography{supplement.bib}